\documentclass[11pt,twoside]{article}
\usepackage{asp2010}

\resetcounters

\begin{document}

\title{The activity and rotation limit in the Hyades}
\author{U. Seemann$^{1,2}$, A. Reiners$^2$, A. Seifahrt$^{2,3}$, and M. K\"urster$^4$
\affil{$^1$European Southern Observatory, Karl-Schwarzschild-Stra\ss e 2, 85748 Garching, Germany; useemann@eso.org}
\affil{$^2$Institut f\"ur Astrophysik, Georg-August-Universit\"at G\"ottingen, Friedrich-Hund-Platz 1, 37077 G\"ottingen, Germany}
\affil{$^3$Department of Physics, University of California, One Shields Avenue, Davis, CA 95616, USA}
\affil{$^4$Max-Planck-Institut f\"ur Astronomie, K\"onigstuhl 17, 69117 Heidelberg, Germany} 
	}

\begin{abstract}
We conduct a study of K to M type stars to investigate the activity and the rotation limit in the Hyades. We use a sample of 40 stars in 
this intermediate-age cluster ($\approx$625 Myr) to probe stellar rotation in the threshold region where stellar activity becomes prevalent. 
Here we present projected equatorial velocities ($v_{rot}\sin i$) and chromospheric activity measurements (H$_\alpha$) that indicate the existence of fast rotators in the 
Hyades at spectral types where also the fraction of stars with H$_\alpha$ emission shows a rapid increase (``H$_\alpha$ limit''). 
The locus of enhanced rotation (and activity) thus seems to be shifted to
earlier types in contrast to what is seen as the rotation limit in field stars. The relation between activity and rotation appears to be
similar to the one observed in fields stars.
\end{abstract}

\section{Introduction}
Solar-type stars are mostly fast rotators and magnetically active when they are young. Their magnetic fields drive stellar winds, which rotationally slow-down the star by means of angular momentum transfer. The stellar spin-down over time is empirically quantified by the so-called ''Skumanich-law`` as $\omega \propto t^{-\frac{1}{2}}$ \citep{Skumanich1972}. 
This relation, however, becomes invalid at very low masses. 
Among the field stars, it is observed that at the transition to fully convective stars at early M-type ($\approx0.3\,M_{\odot}$), the rotational braking efficiency changes, and fast rotation ($v_{\rm{rot}}>3\,$km/s) becomes predominant \citep{Delfosse1998, Mohanty2003, ReinersBasri2008}. The threshold between slow and rapid rotation is thought to be age-dependent \citep{Hawley1999}, so that young cluster stars are expected to show a rotation limit shifted towards higher masses or earlier spectral types, compared to (old) field stars. 

Stellar rotation and magnetic activity are tightly linked by the underlying dynamo processes. 
In (young) clusters, it is observed that the fraction of active stars (eg.\ with chromospheric H$_\alpha$ emission) sharply increases at different masses depending on the cluster age \citep[``H$_\alpha$ limit'',][]{Hawley1999}.
At younger age, enhanced magnetic activity is seen at higher masses (ie.\ earlier spectral types) than it is for older clusters. 
However, it is elusive whether this change in the locus of the H$_\alpha$ limit is also due to an increase of the rapid rotation rate \citep{Radick1987, Stauffer1987}.  Previous studies have focussed on the evolution of the H$_\alpha$ limit in young and intermediate age clusters \citep{Stauffer1997, Hawley1999, Reid1995} or field stars \citep{West2004}, but rotational velocities have only been measured extensively for field stars  all across the main sequence \citep{Delfosse1998, Mohanty2003, West2008, ReinersBasri2008}. However, for open clusters such as the Hyades, $v_{rot}\sin i$ measurements have concentrated on earlier spectral types F to K \citep[eg.][]{Radick1987}, and on the very low-mass regime \citep[M-type and below; eg.][]{ReidMahoney2000}, so that in the mid-K to early M-type range (hence in the range of the H$_\alpha$-limit) rotational velocities are scarce for the Hyades. 

The present work adresses this scarcity between spectral classes K and M, and probes the coexistence of enhanced activity at the H$_\alpha$ limit and rapid rotation for young stars in the case of the intermediate aged Hyades. 
We thus aim to determine if rapid rotation occurs at a different threshold in the Hyades with respect to field stars. 

\section{Sample selection and Observations}
\begin{figure}[!ht]
\plotone{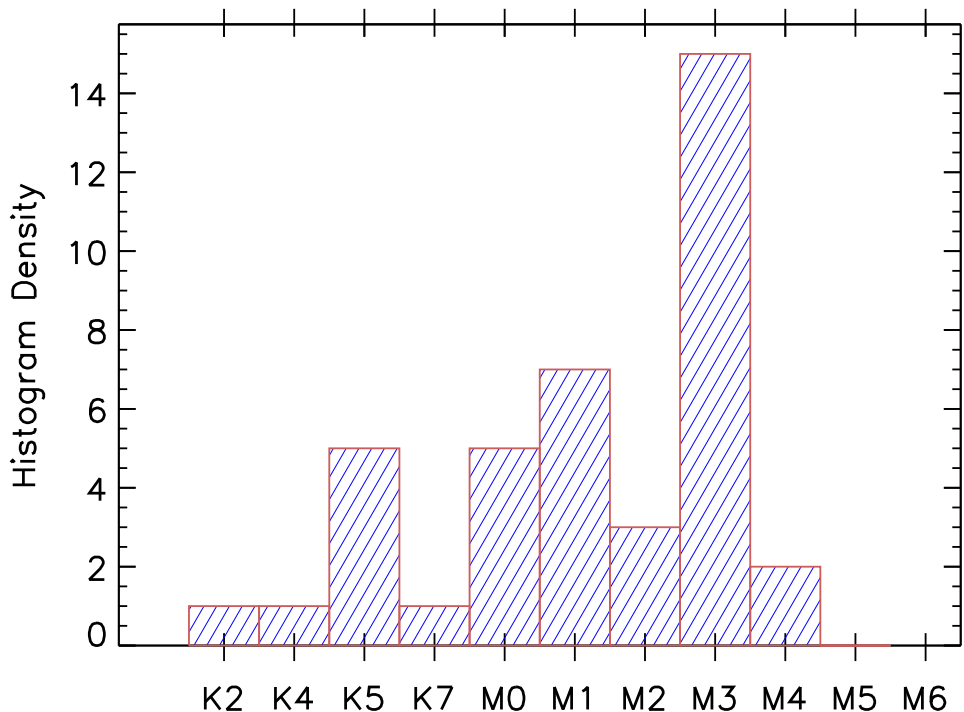}
\caption{
	Distribution of spectral types in the Hyades sample. The total sample size is 40. The K-star bins are two spectral sub-types wide. 
	}
\label{fig_sample} 
\end{figure}
%
Our sample of low-mass stars comprises 40 members of the Hyades open cluster. Selection is based on color, and proper motions where available from the literature. 
All stars are drawn from spectral types early K to mid M, and thus bracket both sides of the H$_\alpha$ limit observed for the Hyades. 
The age of the Hyades open cluster has been estimated as $625\pm50\,$Myr \citep{Perryman1998}, so that also the later M-type members have already settled on the zero age main sequence and are expected to spin down. 

We obtained high-resolution ($R=48\,000$) optical spectra between 360\,nm and 920\,nm for all our sample stars using the Fiber-fed Extended Range Optical Spectrograph (FEROS), mounted on the ESO/MPG 2.2\,m telescope on La Silla. The signal-to-noise ratio exceeds 60 around 800\,nm for the K-type objects. For the M-dwarfs, the SED falls off rapidly towards the blue, so that for these intrinsically faint objects ($V=12..15$) we still achieve a signal-to-noise ratio higher than 30 at 800\,nm. 

\section{Data analysis}
Data reduction of the echelle spectra follows standard procedures employing the FEROS pipeline package within MIDAS. 
From the extracted spectra, H$_\alpha$ ($\lambda656.3\,$nm) equivalent widths are measured as a proxy of chromospheric activity. We express the H$_\alpha$ emission strength relative to the bolometric luminosity, ie.\ $L_{H_\alpha}/L_{\rm{bol}}$ to account for the steeply decreasing luminosity within the spectral range K--M, which would otherwise overestimate H$_\alpha$ emission for earlier spectral types. $L_{\rm{bol}}$ is computed from synthetic Phoenix spectra \citep{Hauschildt1999} of the same spectral type. 

Rotational velocities $v_{rot}\sin i$ are determined by a cross-correlation technique, similar to the methods used by \citet{Browning2010}. 
For bins of adjacent spectral types, template spectra are constructed from slowly rotating stars of very similar spectral type as that of the sample stars. The template stars were also observed with FEROS to minimize the effects of instrumental profile, and have known $v_{rot}\sin i<2.5\,$km/s, which is our detection limit. We consider their rotation as negligible. The templates are then artificially broadened employing a line-broadening kernel, and cross-correlated against the object spectra to construct a line broadening curve, from which $v_{rot}\sin i$ is derived. 
Cross-correlation is performed in about 20 selected wavelength-ranges, typically $0.2-0.5\,$nm wide, between $500-900\,$nm that contain moderately deep, isolated photospheric lines. 
For each object, $v_{rot}\sin i$ is derived from a set of template stars that bracket the object in spectral type.

Spectral classes of the sample objects are determined from the photometric $H-K$ color \citep[2MASS,][]{Skrutskie2006}, which gives a more reliable temperature indicator for the K and M-type stars than optical colors. The $H$ and $K$ magnitudes were also measured simultanously, and hence are free of an activity bias that sequentially taken magnitudes might suffer from. 
The distribution of spectral types covered by the sample is shown in Fig.\,\ref{fig_sample}.

\section{Results}
\begin{figure}[!ht]
\plotone{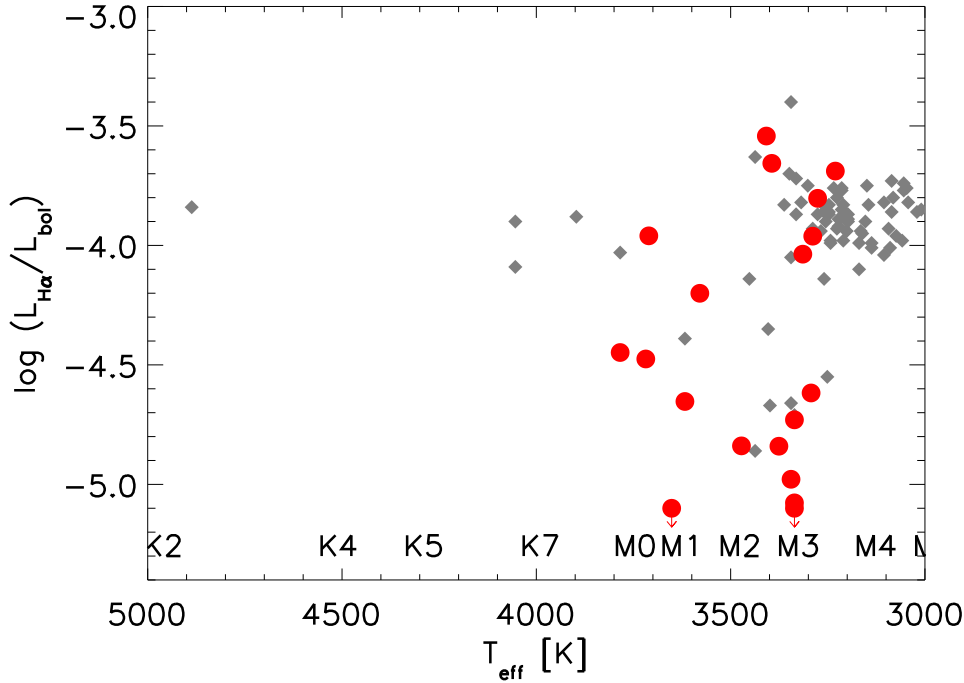}
\caption{
	Normalized H$_\alpha$ activity vs. spectral type for our
	sample Hyads (red points). Data from \citet{ReidMahoney2000} is overplotted as diamonds.
	The combined data indicates that chromospheric magnetic activity (H$_\alpha$ emission) for the Hyades arises at higher masses (M0--M2) than for (old) field stars ($\approx$M3). 
	No predominant activity is seen in K-stars due to magnetic braking (slow rotators, cf.\ Fig.\,\ref{fig_vrot}).
	}
\label{fig_halpha}
\end{figure}
\begin{figure}[!ht]
\plotone{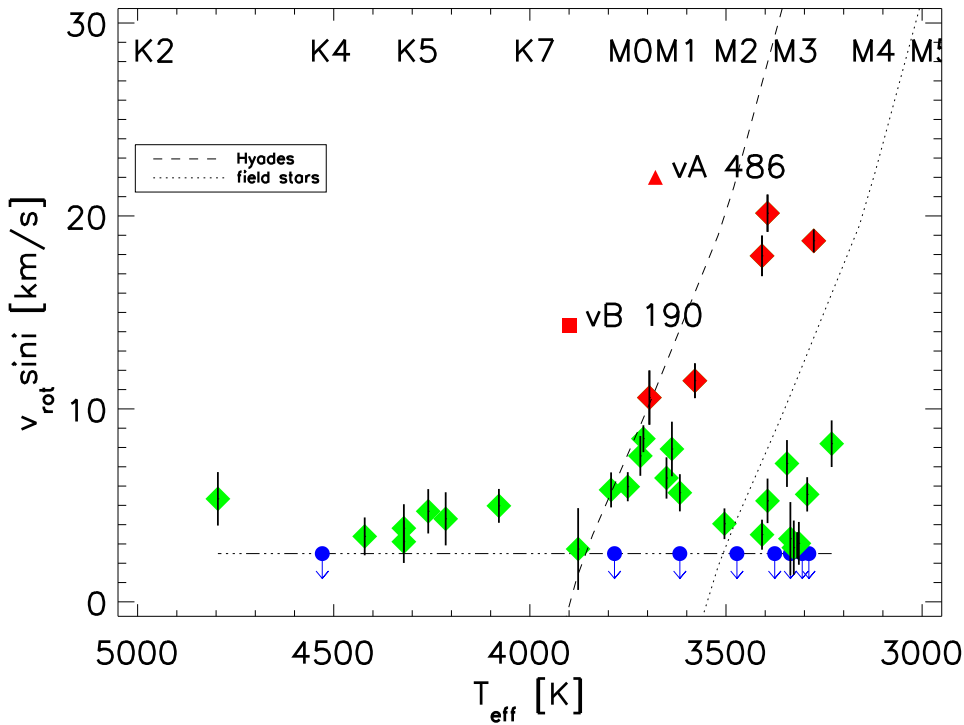}
\caption{
	Measured rotational velocities $v_{rot}\sin i$ as a function
	of temperature (with corresponding spectral types) for the Hyades sample.
	Non-detections (below our detection limit; blue points), rotators
	(green diamonds) and rapid rotators (red diamonds) indicate an increase in
	$v_{rot}\sin i$ towards higher rotation rates in the Hyades around
	$\approx$M0 (dashed line for illustration), in contrast to field stars where this
	threshold kicks in at $\approx$M3 (dotted line). We thus see
	activity and rotation in the 625\,Myr young Hyades at
	earlier spectral types than in old ($\approx$5\,Gyr) field stars.
	VA486 (red triangle) from \citet{Seifahrt2009}, vB190
	(red square) from \citet{Radick1987}.
	}
\label{fig_vrot}
\end{figure}
\begin{figure}[!ht]
\plotone{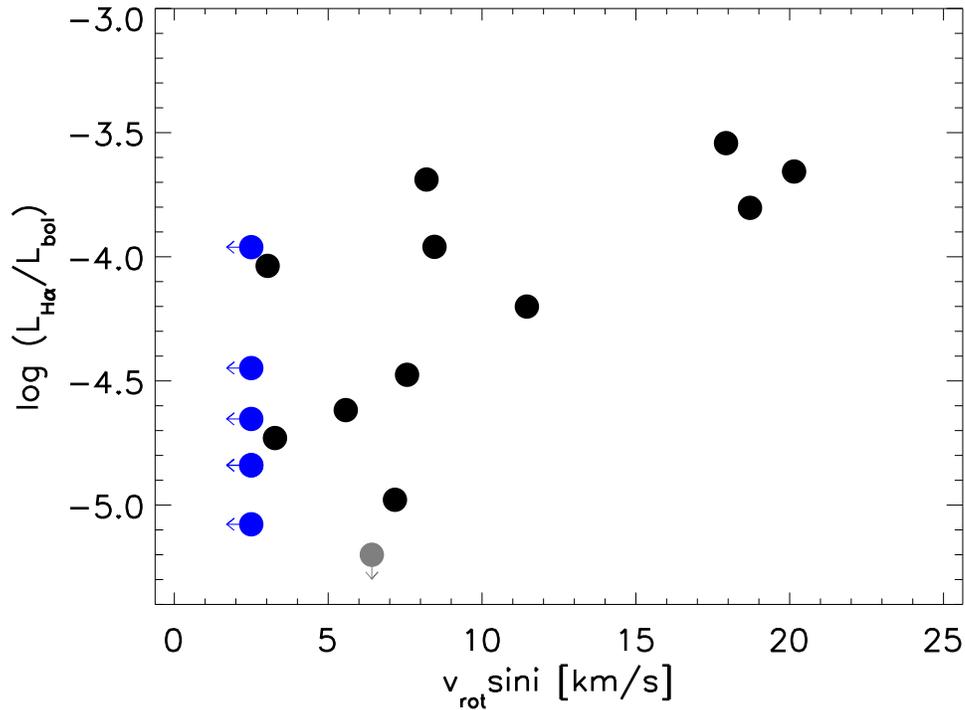}
\caption{
	Normalized H$_\alpha$ activity vs. measured $v_{rot}\sin i$ for
	our sample Hyads (black points; gray and blue points mark
	upper limits in activity and rotational velocity, respectively).
	The most active objects are also the most rapid
	rotators, which is consistent with the rotation-activity
	relation seen in field stars.
	}
\label{fig_actrot}
\end{figure}
\begin{figure}[!ht]
\plotone{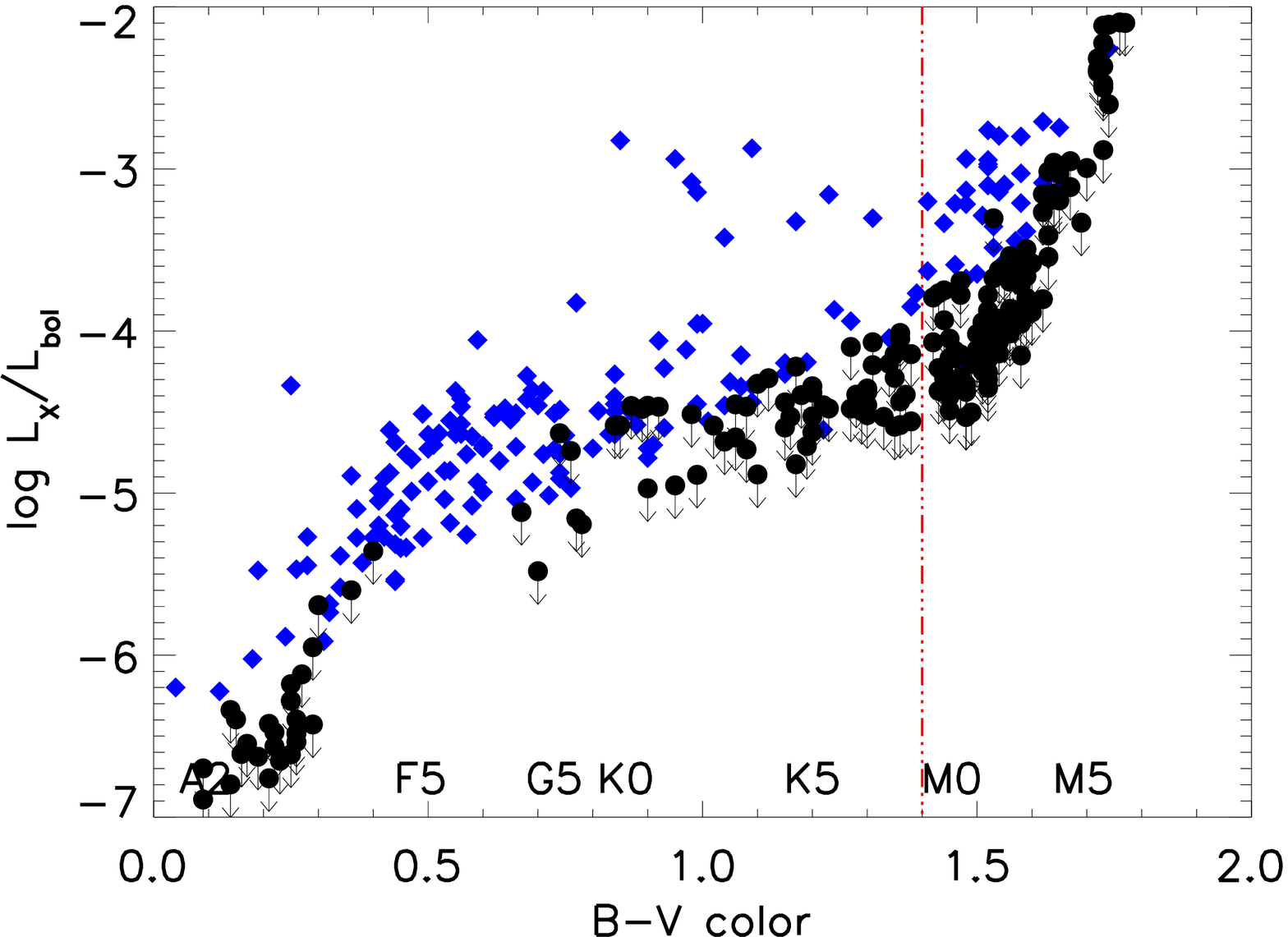}
\caption{
	Normalized X-ray (ROSAT) luminosity as a
	function of (B-V) color (blue diamonds, upper limits as black points; with correspondence to spectral types) shows a strong increase in
	coronal activity at spectral type $\approx$M0 (indicated by dashed line) for the Hyades, analogous to chromospheric activity. 
	Data from the catalog in \citet{Stern1995}.
	}
\label{fig_lx}
\end{figure}
\subsection{Activity}
From the 40 cluster stars, we find significant H$_\alpha$ emission in 18 objects. Their normalized H$_\alpha$ luminosity distribution is plotted in Fig.\,\ref{fig_halpha}. All of the active stars we find within spectral range M0--M4. The strongest emission is detected at spectral type M3, in agreement with data from \citet{ReidMahoney2000}, who studied H$_\alpha$ activity in a large sample of mostly M-type Hyades stars. 
The onset of activity in our data confirms an H$_\alpha$ limit in the Hyades shifted to earlier spectral type when compared to field stars.

\subsection{Rotational velocities}
We detect rotation above the detection limit of $v_{rot}\sin i>2.5\,$km/s in 30 stars, spread over all spectral types in the sample. We find rapid rotation with $v_{rot}\sin i>10\,$km/s among the M dwarfs, with an increase in $v_{rot}\sin i$ towards higher rotation rates around M0, while we do not see any fast rotators ($v_{rot}\sin i>5\,$km/s) among the K stars. The latter is likely due to the stronger rotational braking in the K-type regime. We find that the locus of the rotation limit at $\approx$M0 in the 625\,Myr aged Hyades data is shifted to higher masses, compared to the rotation limit at spectral class $\approx$M3 found in old field stars \citetext{see Joshi et al, these proceedings}. 

Our findings in H$_\alpha$ are in good agreement with previous chromospheric H$_\alpha$ \citep{Hawley1994} and coronal activity measurements from X-rays \citep[Fig.\,\ref{fig_lx};][]{Stern1995, Guedel2004}. In both activity proxies, a threshold in normalized activity luminosity is observed at spectral type $\approx$M0 in the Hyades. The behaviour in $v_{rot}\sin i$ 
seems to coincide, suggesting that the rotation-activity relation known for field stars is still in place in a 625\,Myr cluster (Fig.\,\ref{fig_actrot}). 
This is also supported by recent photometric studies of the Hyades \citetext{see Delorme et al, these proceedings} that show a clear drop-off in rotational period at $J-K_s\approx 0.85$ (spectral type M0). 
\section{Conclusions}
Our sample of Hyads shows increased rotation alongside with H$_\alpha$ activity at spectral types $\approx$M0 and later. We see evidence that young, early M-
type stars rotate faster than field stars do. 
This is possibly influenced by the different contraction timescales present in these young stars over this range of spectral types, giving rise to different braking efficiencies and histories.

\acknowledgements We thank Ulli K\"aufl and Nandan Joshi for helpful discussions and advice. US and AR acknowledge research funding
from the DFG under grant RE\,1664/4-1. This research has made use of the Simbad and Vizier databases, operated at CDS, Strasbourg, France, and NASA's Astrophysis Data System Bibliographic Services. 

\bibliography{seemann_u}

\end{document}